\documentclass[12pt]{iopart}
\usepackage{amssymb}
\usepackage{graphicx}

\begin{document}

\title{Geometrical aspects and connections of the energy-temperature fluctuation relation}
\author{L. Velazquez$^{1,2}$ and S. Curilef$^{2}$}

\begin{abstract}
Recently, we have derived a generalization of the known canonical
fluctuation relation $k_{B}C=\beta^{2}\left\langle \delta
U^{2}\right\rangle $ between heat capacity $C$ and energy
fluctuations, which can account for the existence of macrostates
with negative heat capacities $C<0$. In this work, we presented a
panoramic overview of direct implications and connections of this
fluctuation theorem with other developments of statistical
mechanics, such as the extension of canonical Monte Carlo methods,
the geometric formulations of fluctuation theory and the relevance
of a geometric extension of the Gibbs canonical ensemble that has
been recently proposed in the literature.\newline
\end{abstract}

\address{$^1$Departamento de F\'\i sica, Universidad de Pinar del R\'\i o,
Mart\'\i \,\,270, Esq. 27 de Noviembre, Pinar del R\'\i o, Cuba.
\\$^2$Departamento de F\'\i sica, Universidad Cat\'olica del Norte, Av. Angamos 0610, Antofagasta,
Chile.}

\section{Introduction}

Recently, we have obtained a suitable extension of the canonical
fluctuation-dissipation relation involving the heat capacity $C$ and
energy fluctuations \cite{vel-unc}:
\begin{equation}
k_{B}C=\beta ^{2}\left\langle \delta U^{2}\right\rangle
+C\left\langle \delta \beta _{\omega }\delta U\right\rangle ,
\label{fdr}
\end{equation}%
which considers a \textit{system-surroundings} equilibrium situation
in which the inverse temperature $\beta _{\omega }=1/T_{\omega}$ of
a given thermostat exhibits non-vanishing correlated fluctuations
with the total energy $U$ of the system under study as a consequence
of the underlying thermodynamic interaction, $\left\langle \delta
\beta _{\omega }\delta U\right\rangle \not=0$. Clearly,
Eq.(\ref{fdr}) differs from the canonical equilibrium situation due
to the realistic possibility that the internal thermodynamical state
of the thermostat can be affected by the presence of the system
under study. This allows us to describe the fluctuating behavior of
the system under more general equilibrium situations, rather than
the ones associated with the known canonical and microcanonical
ensembles.

The fluctuation relation (\ref{fdr}) possesses interesting
connections with some challenging problems related to statistical
mechanics, such as: (1) a compatibility with the existence of
macrostates exhibiting \textit{negative heat capacities} $C<0$
\cite{vel-unc,vel-stab}, a thermodynamic anomaly that appears in
many physical contexts (ranging from small nuclear, atomic and
molecular clusters
\cite{moretto,Ison,Dagostino,gro na} to the astrophysical systems \cite%
{Lynden,Lyn2,thir,Einarsson}) associated with the existence of
\textit{\ nonextensive properties} \cite{pad,Lyn3,gro1,Dauxois}; (2)
a direct application for the extension of available Monte Carlo
methods based on the consideration of the Gibbs canonical ensemble
in order to capture the presence of a regime with $C<0$ and avoid
the incidence of the so-called \textit{super-critical slowing down}
\cite{vel-unc,vel-mc} (a dynamical anomaly associated with the
occurrence of discontinuous (first-order) phase transitions
\cite{Reichl}, which significantly reduces the efficiency of Monte
Carlo methods \cite{mc3}); (3) finally, a direct relationship with
an \textit{uncertainty relation} supporting the existence of some \textit{%
complementary character} between thermodynamic quantities of energy
and temperature \cite{vel-unc,vel-complem}, an idea previously
postulated by Bohr and Heisenberg \cite{bohr,Heisenberg} with a long
history in the literature
\cite{Rosenfeld,Scholgl,Mandelbrot,Uffink}.

Our aim in this work is to present a more complete study of the
existing connections of the fluctuation-dissipation relation
(\ref{fdr}). The core of our analysis is focussed on certain
geometric aspects relating the present approach with other geometric
formulations of fluctuation theory \cite{rupper}. Such ideas
straightforwardly lead to a geometric generalization of the Gibbs
canonical ensemble describing a special family of equilibrium
distributions recently proposed in the literature
\cite{hugo.gce,toral}, which can also be obtained from some known
formulations of statistical mechanics, such as Jaynes'
reinterpretation in terms of information theory \cite{jaynes}, as
well as Mandelbrot's approach based on inference theory
\cite{mandelbrot.it}.

\section{A brief review}

\subsection{Compatibility with negative heat capacities}

Our main motivation in deriving the fluctuation-dissipation relation (\ref%
{fdr}) was to arrive at a suitable extension of the known
fluctuation relation:
\begin{equation}
k_{B}C=\beta ^{2}\left\langle \delta U^{2}\right\rangle  \label{can}
\end{equation}%
that is compatible with the existence of macrostates with negative heat
capacities \cite{vel-unc,vel-stab}. As discussed in many standard
textbooks on statistical mechanics \cite{Reichl}, the latter
relation follows as a direct consequence of the consideration of the
Gibbs canonical ensemble:
\begin{equation}
p_{c}\left( \left. U\right\vert \beta \right) =\frac{1}{Z\left(
\beta \right) }\exp \left( -\frac{1}{k_{B}}\beta U\right) \Omega
\left( U\right) dU,  \label{gibbs}
\end{equation}%
which constitutes a starting point for many applications of
equilibrium statistical mechanics. However, such a relation is only
compatible with macrostates having non-negative heat capacities, and
hence, all those macrostates with negative heat capacities $C<0$
cannot be appropriately described by using this statistical
ensemble. In fact, such macrostates are thermodynamically unstable
under this kind of equilibrium situations (a system submerged in a
certain environment (heat reservoir or bath) with constant inverse
temperature $\beta $).

One can easily verify from Eq.(\ref{fdr}) that a macrostate with a
negative heat capacity $C<0$ is thermodynamically stable provided
that the correlation function $\left\langle \delta \beta _{\omega
}\delta U\right\rangle $ considering the existence of correlative
effects between the system and its surroundings obeyed the following
inequality:
\begin{equation}
\left\langle \delta \beta _{\omega }\delta U\right\rangle >k_{B}.
\label{cc}
\end{equation}

A simple interpretation (but not the only one possible) of the above
fluctuating constraint follows from admitting that the thermostat or
the surroundings is a finite system with a positive heat
capacity $C_{\omega }$. Clearly, the existing energetic interchange
between these systems imposes the occurrence of thermal fluctuations
for the thermostat temperature $T_{\omega }$, $\delta T_{\omega
}=-\delta U/C_{\omega }$, with $\delta U$ the amount of energy
released or absorbed by the system around its equilibrium value.
Such fluctuations can be rephrased as follows:
\begin{equation}
\delta \beta _{\omega }=\beta ^{2}\delta U/C_{\omega },  \label{bb}
\end{equation}%
where the condition of thermal equilibrium $\beta =\beta _{\omega }$
is considered. By substituting Eq.(\ref{bb}) into the
fluctuation-dissipation
relation (\ref{fdr}), we obtain:%
\begin{equation}
k_{B}\frac{CC_{\omega }}{C+C_{\omega }}=\beta ^{2}\left\langle
\delta U^{2}\right\rangle .  \label{bp}
\end{equation}%
Finally, it is possible to arrive at the following inequalities:
\begin{equation}
\frac{C}{C+C_{\omega }}>1\Leftrightarrow 0<C_{\omega }<\left\vert
C\right\vert   \label{c.thir}
\end{equation}%
by combining Eqs.(\ref{cc})-(\ref{bp}). Essentially, this last result
is the same constraint derived by Thirring in order to ensure the
thermodynamic stability of macrostates with a negative heat capacity
\cite{thir}.

\subsection{Extension of canonical Monte Carlo methods}

\label{Monte}

The study of macrostates with negative heat capacities demands that
such macrostates be found in a stable equilibrium situation. As
already discussed, such an aim could be implemented by considering
an equilibrium situation in which the system is found in thermal
contact with a bath having a \textit{positive and finite heat
capacity} $C_{\omega }$\ that obeys Thirring's constraint
(\ref{c.thir}). The equilibrium condition associated with the Gibbs
canonical ensemble (\ref{gibbs}) is unsuitable here, since the
invariability of the Gibbs thermostat temperature presupposes a
system with an infinite heat capacity, $C_{\omega }\rightarrow
+\infty $, which is incompatible with inequality (\ref{c.thir}).

\begin{figure}[t]
\begin{center}
\includegraphics[
height=2.725in, width=3.5397in ]{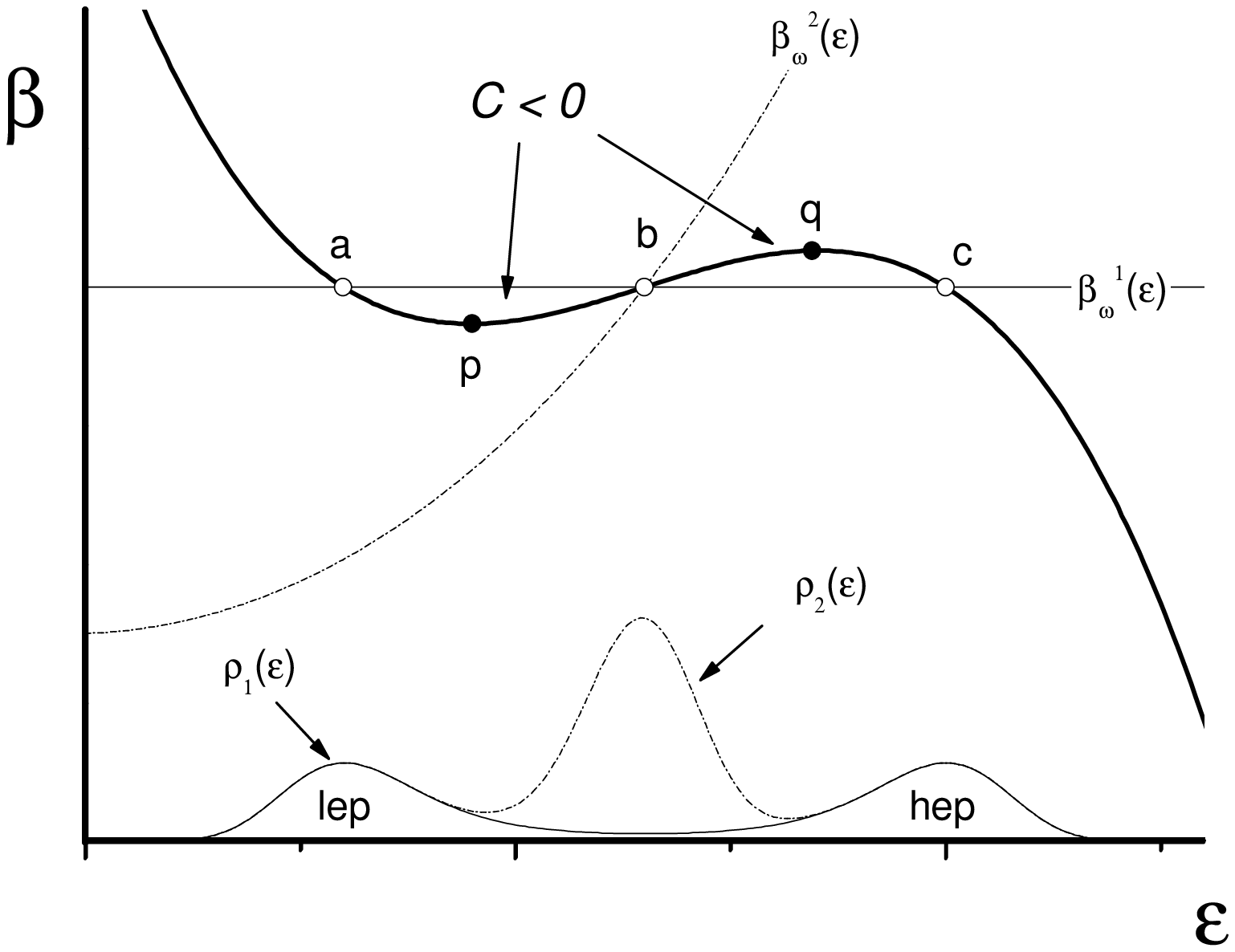}
\end{center}
\caption{Schematic behavior of the microcanonical caloric $\protect\beta %
\left( \protect\varepsilon \right) =\partial s\left( \protect\varepsilon %
\right) /\partial \protect\varepsilon $ of a finite short-range
interacting
system undergoing a first-order phase transition. Here, $\protect\rho %
_{1}\left( \protect\varepsilon \right) $ and $\protect\rho
_{2}\left( \protect\varepsilon \right) $ respectively represent the
energy distribution functions when this system is placed in thermal
contact with a Gibbs
thermostat with inverse temperature $\protect\beta _{\protect\omega%
}^{1}\left( \protect\varepsilon \right) =const$ and a heat bath
having a finite positive heat capacity and, therefore, a variable
(fluctuating)
inverse temperature $\protect\beta _{\protect\omega}^{2}\left( \protect%
\varepsilon \right) $.} \label{caricature.eps}
\end{figure}

The differences between these equilibrium situations are
schematically illustrated in FIG.\ref{caricature.eps}. Here, the
thick solid line
represents the typical microcanonical caloric curve $\beta\left(\varepsilon%
\right)=\partial s\left(\varepsilon\right)/\partial \varepsilon$ of
a finite short-range interacting system undergoing a first-order
phase transition, which is characterized by the existence of a
regime with negative heat capacities (the branch p-q), with
$\varepsilon=U/N$ being the energy per particle. The thin solid lines
$\beta^{1}_{\omega} \left(\varepsilon\right)$ and
$\beta^{2}_{\omega}\left(\varepsilon\right)$ are respectively the
inverse temperature dependencies on the system energy per particle $%
\varepsilon$ of a Gibbs thermostat (with
$C_{\omega}\rightarrow+\infty$) and a thermostat having a positive
finite heat capacity $0<C_{\omega}<+\infty
$, with $\rho_{1}\left(\varepsilon\right)$ and $\rho_{2}\left(\varepsilon%
\right)$ being the corresponding energy distribution functions.

The intersection points derived from the condition of thermal equilibrium $%
\beta \left( \varepsilon \right) =\beta _{\omega }\left( \varepsilon
\right) $ determine the positions of the energy distribution function
$\rho \left( \varepsilon \right)$ maxima and minima.
Clearly, the thermal contact with a Gibbs thermostat ensures the
existence of only one intersection point, or equivalently, a unique
peak of the canonical energy distribution function $\rho _{1}\left(
\varepsilon \right) $ for most of the admissible values of the
thermostat inverse temperature. The important
exception takes place in the inverse temperature interval $\left[ \beta _{p},\beta _{q}%
\right] $, where there are \textit{three intersection points} (two maxima $%
\varepsilon _{a}$ and $\varepsilon _{c}$ with $C>0$ and one minimum $%
\varepsilon _{b}$ with $C<0$), a fact that leads to a
\textit{bimodal character} for the distribution function $\rho
_{1}\left( \varepsilon \right) $. Since no single peak can be located
within the branch with negative heat capacities $C<0$, such
macrostates are poorly accessed within the Gibbs canonical ensemble.
In fact, they turn practically inaccessible when the system size is sufficiently
large. The existence of such a \textit{hidden} energetic
region constitutes the origin of the latent heat $q_{L}$ necessary
for the conversion of one phase into the other during the
coexistence of low and high energy phases (lep and hep), which
are represented here by the coexisting peaks of the canonical energy
distribution function $\rho _{1}\left( \varepsilon \right) $.

The replacement of the Gibbs thermostat by a thermostat having a
finite positive heat capacity crucially modifies the fluctuating
behavior and the thermal stability conditions of the system. In
fact, one can ensure the existence of only one intersection point,
regardless the positive or negative character of its heat capacity
$C$, by choosing the appropriate thermostat and its internal
conditions. In particular, it is necessary to ensure the applicability
of Thirring's constraint (\ref{c.thir}) for macrostates with
negative heat capacities $C<0$.

The above ideas have a significant impact in the framework of
Monte Carlo simulations. As has been discussed elsewhere \cite{mc3},
large-scale Monte Carlo simulations are often plagued by slow
sampling problems, which manifest themselves as a rapid increase in
the dynamic relaxation time $\tau$ with the system size $N$, causing
large-size simulations to converge extremely slowly. These sampling
problems are especially severe in systems near the critical point,
where it is possible to distinguish two kinds of dynamical
anomalies: (1) the so-called \textit{critical slowing down}, where
the relaxation time shows a power-law dependence on $N$,
$\tau\propto N^{\alpha}$, which can be associated with the occurrence
of a continuous (second-order) phase transition; and (2) the
so-called \textit{super-critical slowing down}, where the dynamic
relaxation time exhibits a worse divergence with the system size: an
exponential increasing $\tau\propto \exp\left(\alpha N\right)$,
whose incidence is associated with discontinuous (first-order) phase
transitions.

In Monte Carlo simulations based on a consideration of the Gibbs
canonical ensemble, the origin of the super-critical slowing down is
closely related to the existence of a multimodal character of the
energy distribution function. Indeed, this phenomenon manifests itself as
an \textit{effective trapping} of the system macrostates in one of
the coexisting peaks of the energy distribution function. As the system
size increases, the mathematical form of these peaks is almost
a Gaussian distribution:
\begin{equation}
\rho\left(\varepsilon\right)\simeq\frac{1}{\sqrt{2\pi\sigma^{2}}}\exp\left[%
-\left(\varepsilon -
\bar{\varepsilon}\right)^{2}/2\sigma^{2}\right],
\end{equation}
whose width behaves as $\sigma\propto 1/\sqrt{N}$. The transition to
any other peak demands the occurrence of a large energy fluctuation,
whose probability $p$ exponentially decreases as the system size
decreases: $p\propto \exp\left(-\alpha N\right)$.
Consequently, the characteristic timescale for the occurrence of
such rare events grows as $\tau\propto 1/p\sim\exp\left(\alpha
N\right)$, which explains the slow relaxation observed for
canonical expectation values in large-scale Monte Carlo simulations.
The existence of the above slow relaxation can be avoided if one could
eliminate the multimodal character of the energy distribution
function by appealing to a better control of the energy
fluctuations. Fortunately, \textit{such an aim is easily achieved by
considering a thermostat having a finite positive heat capacity}
$C_{\omega}$.

Under this later equilibrium situation, the thermostat inverse temperature $%
\beta_{\omega}$ and the system energy (per particle) $\varepsilon$
undergo
thermal fluctuations around their equilibrium values $\left\langle\beta_{%
\omega}\right\rangle$ and $\left\langle\varepsilon\right\rangle$,
which provide a suitable estimation of the intersection point of the
system microcanonical caloric curve
$\beta\left(\varepsilon\right)=\partial
s\left(\varepsilon\right)/\partial\varepsilon$ derived from the
thermal
equilibrium condition $\beta\left(\varepsilon\right)=\beta_{\omega}\left(%
\varepsilon\right)$. Moreover, the study of the fluctuating behavior
in terms of
correlation functions $\left\langle\delta\beta_{\omega}\delta\varepsilon%
\right\rangle$ and $\left\langle\delta\varepsilon^{2}\right\rangle$
allows us to
obtain the heat capacity $C$ via the fluctuation-dissipation relation (\ref%
{fdr}). Once the microcanonical caloric curve $\beta\left(
\varepsilon\right) =\partial s\left( \varepsilon\right)
/\partial\varepsilon$ has been obtained, one can easily derive other
thermodynamic potentials by using known integration formulae,
e.g.: the entropy $s\left(
\varepsilon\right) $:%
\begin{equation}
\Delta s\left( \varepsilon\right) =s\left( \varepsilon\right)
-s\left( \varepsilon_{0}\right)
=\int_{\varepsilon_{0}}^{\varepsilon}\beta\left( \varepsilon\right)
d\varepsilon,   \label{ss}
\end{equation}
the Helmholtz free energy $f\left( \beta\right) =-T\log Z\left(
\beta\right)
/N$:%
\begin{equation}
Z\left( \beta\right) =N\int\exp\left\{ -N\left[
\beta\varepsilon-s\left( \varepsilon\right) \right] \right\}
d\varepsilon,
\end{equation}
and the canonical averages of a certain observable $O\left(\varepsilon\right)$%
:
\begin{equation}
\left\langle O\right\rangle=\frac{1}{Z\left( \beta\right)}N\int
O\left(\varepsilon\right)\exp\left\{ -N\left[
\beta\varepsilon-s\left( \varepsilon\right) \right] \right\}
d\varepsilon.
\end{equation}

\begin{figure}[t]
\begin{center}
\includegraphics[
height=2.7in, width=3.4in ]{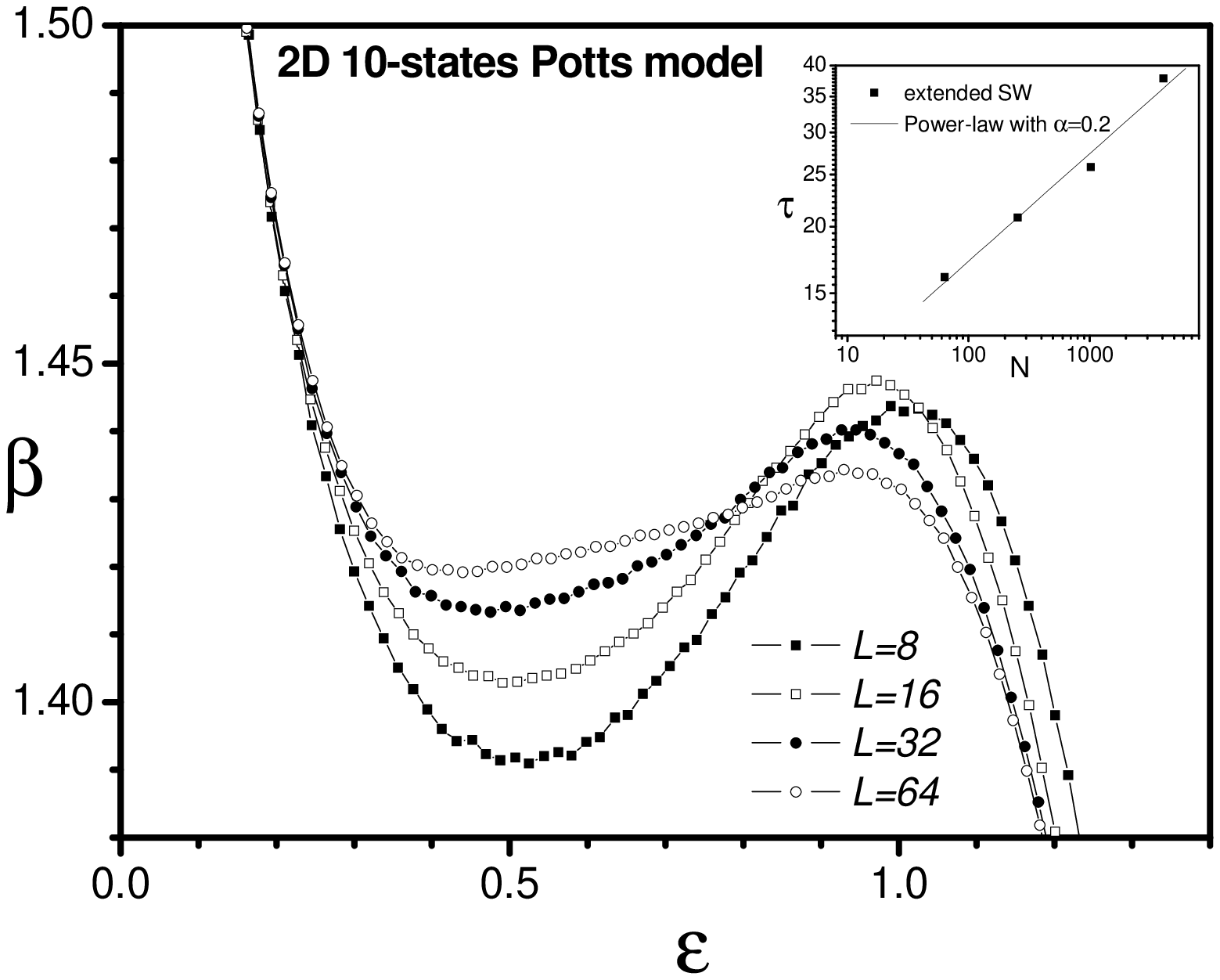}
\end{center}
\caption{Microcanonical caloric curves of the $2D$ $10$-states Potts
model on a square lattice $L\times L$ obtained from Monte Carlo
simulations using the extended version of the Swendsen-Wang clusters
algorithm (extended SW). Inset panel: Power-law dependence of the
decorrelation time $\tau$ with the system size $N=L^{2}$ at the
critical point $\beta_{c}$, $\tau\sim N^{\alpha}$, with
$\alpha\simeq 0.2$. } \label{montecarlo.eps}
\end{figure}

The simplest and most general way to implement the use of a
thermostat having a finite heat capacity in a classical Monte Carlo
calculation is through the known Metropolis importance sample
\cite{met}. Its extension is achieved by replacing the use of a
constant inverse temperature in the acceptance probability:
\begin{equation}
p\left( U|U+\triangle U\right) =\min \left\{ \exp \left( -\beta
_{B}\Delta U\right) ,1\right\}
\end{equation}%
with a variable inverse temperature, $\beta _{B}\rightarrow \beta
_{\omega }\left( \varepsilon \right) $. This kind of procedure can
also be used to extend some other classical Monte Carlo methods,
such as the known Swendsen-Wang (SW) clusters algorithm
\cite{SW,pottsm,wolf}, applicable to the Ising model and its
generalization, the $q$-state Potts model:
\begin{equation}
H_{q}=\sum_{ij\in n.n}\left( 1-\delta _{\sigma _{i},\sigma
_{i}}\right)
\end{equation}%
(where n.n represents a set of nearest-neighbor lattice sites, $\sigma _{i}=[1,2,\ldots
,q]$), which exhibits a regime with negative heat capacities when
the number of spin states $q$ is greater than a certain critical
value depending on the lattice dimensionality $D$, e.g. $q>3$ with
D=2. A direct demonstration of the applicability of the extended SW
method using the present ideas in order to study the anomalous
regime with $C<0$ in the $2D$ $10$-state Potts model is shown in
FIG.\ref{montecarlo.eps}, whose decorrelation time $\tau$ shows a
weak power-law dependence $\tau\sim N^{\alpha}$ with $\alpha\simeq
0.2$ at the critical point of the discontinuous phase transition
$\beta_{c}$.

Generally speaking, the consideration of a finite thermostat in
order to capture the anomalous regime with negative heat capacities
and to avoid the super-critical slowing down should not depend on the
classical or quantum nature of the system under analysis.
Consequently, one can expect that this idea could be used for
enhancing the potentialities of some known quantum Monte Carlo
methods.

\subsection{Complementarity character between energy and temperature}

\label{comple}

The fluctuation-dissipation relation (\ref{fdr}) constitutes a
particular case of a very general fluctuation relation:
\begin{equation}  \label{mother}
\left\langle \delta \eta \delta U\right\rangle =k_{B}
\end{equation}%
involving the inverse temperature difference between the
surroundings (heat reservoir or bath) and the system $\eta =\beta
_{\omega }-\beta $. In fact, Eq.(\ref{fdr}) is obtained after
substituting the first-order approximation:
\begin{equation}  \label{g.c.fdr}
\delta\beta\simeq -\beta^{2}\delta U/C
\end{equation}
into Eq.(\ref{mother}).

Alternatively, one can consider the known \textit{Schwartz
inequality}:
\begin{equation}
\left\langle \delta A \delta B\right\rangle ^{2}\leq\left\langle
\delta A^{2} \right\rangle \left\langle \delta B^{2} \right\rangle
\end{equation}%
in order to rewrite the fluctuation relation (\ref{mother}) as
follows:
\begin{equation}  \label{unc}
\Delta \eta\Delta U\geq k_{B},
\end{equation}
where $\Delta x=\sqrt{\left\langle\delta x^{2} \right\rangle}$
denotes the thermal uncertainty of a physical observable $x$.
Clearly, Eq.(\ref{unc}) is a thermo-statistic analogy of the
quantum mechanics uncertainty relation:
\begin{equation}  \label{uncqp}
\Delta q\Delta p\geq \hbar
\end{equation}
between position $q$ and momentum $p$, which suggests the existence
of certain complementary character between thermodynamic quantities
of energy
and (inverse) temperature \cite%
{bohr,Heisenberg,Rosenfeld,Scholgl,Mandelbrot,Uffink}.

It is well-known that the nature of the temperature is radically
different from a direct observable quantity such as energy. In fact,
it is a thermodynamic quantity whose physical meaning can only be
attributed by appealing to the concept of statistical ensemble. In
practice, the system temperature is indirectly measured by using the
temperature of a second system through the thermal equilibrium
condition, which plays the role of a measuring apparatus
(thermometer), whose internal temperature dependence on some direct
\textit{thermometric quantity} (e.g.: electric signal, force,
volume, etc.) is previously known. As expected, such a measuring
process unavoidably involves a perturbation on the internal state of
the system under analysis.

According to uncertainty relation (\ref{unc}), it is impossible to
simultaneously reduce the thermal
uncertainties of the inverse temperature difference $\Delta \eta$
and the system energy $\Delta U$ to zero: any attempt to reduce the
perturbation of the system energy to zero, $\Delta U\rightarrow 0$,
leads to a divergence of the inverse temperature difference
uncertainty $\Delta \eta\rightarrow\infty$, and vice-versa.
Consequently, \textit{it is impossible to simultaneously determine
 of the energy and inverse temperature of a given system
using the standard experimental procedures based on the thermal
equilibrium with a second system}%
.

Clearly, we have to admit non-vanishing thermal uncertainties
$\Delta \eta$ and $\Delta U$ during any practical determination of
the energy-temperature dependence of a given system, that is, its
caloric curve. While such thermal uncertainties are unimportant
during the study of large thermodynamic systems, they actually
impose a fundamental limitation to the practical utility of
thermodynamic concepts such as temperature and heat capacity in
systems with few constituents. In order to avoid any
misunderstanding, it must be clarified that one can obtain the
energy dependence of the inverse temperature of a given system
\textit{by calculating} its Boltzmann's entropy:
\begin{equation}  \label{boltzmann}
S=k_{B}\log W\rightarrow \beta=1/T=\partial S/\partial U,
\end{equation}
which is possible to be achieved regardless of the system size. The
limitation associated with the uncertainty relation (\ref{unc})
refers to the precision of an \textit{experimental measuring} of the
microcanonical caloric curve of a thermodynamic system.

\section{Geometrical aspects in fluctuation theory}

\subsection{Starting considerations}

As previously discussed in detail in our first paper on this subject \cite%
{vel-unc}, the rigorous fluctuation relation (\ref{mother}) is derived
from the following ansatz for the energy distribution function:
\begin{equation}  \label{ansatz}
dp=\rho\left(U\right)dU=\omega\left(U\right)\Omega\left(U\right)dU,
\end{equation}
where $\Omega\left(U\right)$ is the state density of the system and $%
\omega\left(U\right)$ is the probabilistic weight considering the
thermodynamic influence of the surroundings (thermostat). Such
functions
are defined on a certain subset $M_{u}$ of Euclidean real space $R$, $%
M_{u}\subset R: U\in\left[U_{inf},U_{sup}\right]$.

The next important consideration is the definition of the
\textit{effective} inverse temperature of the surroundings as:
\begin{equation}  \label{beta.w}
\beta_{\omega}\left(U\right)=-k_{B}\frac{\partial \log\omega\left(U\right)}{%
\partial U}.
\end{equation}
The latter assumption is not arbitrary, since it reduces to the
conventional interpretation of this concept when one considers a
closed system composed of two separable short-range interacting
systems in thermal contact and a final thermodynamic equilibrium,
which allows us to express the probabilistic weight
$\omega\left(U\right)$ in terms of the state density of the second
system $\Omega_{B}\left(U_{B}\right)$, $\omega\left(U\right)\propto%
\Omega_{B}\left(U_{T}-U\right)$.

However, the probabilistic weight $\omega\left(U\right)$ in Eq.(\ref{ansatz}%
) also admits more general system-surroundings equilibrium
situations considering other modifying conditions, such as the
existence of nonlinear effects driving the system-surroundings
thermodynamic interaction, e.g., the presence of long-range
interactions \cite{inChava,beck1}, or a system acting as the
surroundings that is found in a metastable equilibrium whose
relaxation time is so long that its dynamic evolution
can practically be disregarded, such as the case of systems with \textit{glassy dynamics} \cite%
{leticia}. Such circumstances explain why we refer to the inverse
temperature (\ref{beta.w}) as \textit{effective}.

The number of microstates $W$ used to obtain the Boltzmann's entropy
(\ref{boltzmann}) is given by the \textit{coarsed grained}
definition:
\begin{equation}  \label{coarse}
W=\Omega \delta\epsilon,
\end{equation}
with $\delta\epsilon$ being a certain small constant energy that
makes $W$
dimensionless. The work hypothesis (\ref{ansatz}) and definitions (\ref%
{beta.w}) and (\ref{coarse}) allow us to express the inverse
temperature difference as:
\begin{equation}
\eta\left(U\right)=\beta_{\omega}\left(U\right)-\beta\left(U\right)%
\equiv-k_{B}\frac{\partial \log\rho\left(U\right)}{\partial U}.
\end{equation}
With the above relation, one can easily obtain the thermodynamic
identities:
\begin{equation}  \label{eq.cond}
\left\langle\eta\right\rangle=\int^{U_{sup}}_{U_{inf}}\eta\left(U\right)\rho%
\left(U\right)dU=0,
\end{equation}
\begin{equation}  \label{aux.fu}
\left\langle
U\eta\right\rangle=\int^{U_{sup}}_{U_{inf}}U\eta\left(U\right)\rho\left(U%
\right)dU=k_{B},
\end{equation}
which are derived by integrating by parts and considering the
following boundary conditions:
\begin{equation}  \label{bu1}
\rho\left(U_{inf}\right)=\rho\left(U_{sup}\right)=0
\end{equation}
\begin{equation}  \label{bu2}
\frac{\partial\rho\left(U_{inf}\right)}{\partial U}=\frac{%
\partial\rho\left(U_{sup}\right)}{\partial U}=0.
\end{equation}

Eq.(\ref{eq.cond}) is simply the thermal equilibrium condition
expressed in terms of \textit{statistical expectation values}:
\begin{equation}
\left\langle\beta\right\rangle=\left\langle\beta_{\omega}\right\rangle.
\end{equation}
This rigorous result clarifies that the known equalization of
(inverse) temperatures during the thermodynamic equilibrium of two
systems actually takes place in an average sense. The fluctuation
relation of Eq.(\ref{mother}) is obtained from Eqs.(\ref{eq.cond})
and (\ref{aux.fu}) after using the identity $\left\langle\delta
U\delta\eta\right\rangle=\left\langle
U\eta\right\rangle-\left\langle U\right\rangle\left\langle \eta\right\rangle$%
.

Let $A$ be a continuous and differentiable function on $M_{u}$,
which also admits a bound expectation value $\left\langle
A\right\rangle$, $ \left|\left\langle
A\right\rangle\right|<+\infty$. Under these assumptions, one can
obtain the following thermodynamic identity:
\begin{equation}  \label{gen.identity}
\left\langle k_{B}\frac{\partial A}{\partial U}
\right\rangle=\left\langle A\eta\right\rangle\equiv\left\langle
\delta A \delta\eta\right\rangle.
\end{equation}
In particular, this  identity reduces to Eq.(\ref{eq.cond}) and
Eq.(\ref{aux.fu}) for $A=1$ and $A=U$ respectively. Moreover, it
also drops to the remarkable fluctuation relation:
\begin{equation}  \label{comple.flu}
\left\langle -k_{B}\frac{\partial\eta}{\partial U}+\delta\eta^{2}\right%
\rangle=0
\end{equation}
for $A=\eta$. This latter identity, hereafter referred to as the
\textit{complementary fluctuation relation}, accounts for the same
information about the system stability conditions derived from the
fluctuation relation of Eq.(\ref{mother}).
For instance, by using the Gaussian approximation (see subsection \ref%
{Gaussian.sub} below):
\begin{equation}
\left\langle \frac{\partial\eta\left(U\right)}{\partial U}\right\rangle\simeq%
\frac{\partial\eta\left( \bar{U}\right)}{\partial U}
\end{equation}
and focusing on the equilibrium situation between two separable
short-range interacting systems, we obtain the fluctuation relation:
\begin{equation}
\beta^{2}\frac{C+C_{\omega}}{CC_{\omega}}k_{B}=\left\langle\delta\eta^{2}%
\right\rangle,
\end{equation}
which leads to the same stability criterion derived from
Eq.(\ref{bp}).

\subsection{Reparametrization invariance}

Let us consider a bijective application $\Theta\left(U\right):
R\rightarrow R $, which is a piece-wise continuous and two time
differentiable function of the variable $U$. Such a function allows
for the existence of a bijective map $\Theta: M_{u}\rightarrow
M_{\phi}$ of
the subset $M_{u}$ on another subset $M_{\phi}\subset R: \Theta\in\left[%
\Theta_{inf},\Theta_{sup}\right]$. It could be said that these
subsets constitute two equivalent \textit{coordinate
representations} of all admissible macrostates of the system, which
shall be denoted as $R_{u}$ and $R_{\phi}$ respectively. The
coordinate transformation induced by the bijective function
$\Theta\left(U\right)$ is referred to as a \textit{reparametrization}.

Since the elementary subset $\left[U,U+dU\right]$ represents the
same system
macrostates considered by the elementary subset $\left[\Theta,\Theta+d\Theta%
\right]$, the elementary probability $dp$ that the system is found
in such conditions, Eq.(\ref{ansatz}), does not depend on the
coordinate representation used for its expression:
\begin{equation}  \label{ele.prob}
dp=\rho_{u}\left(U\right)dU=\rho_{\phi}\left(\Theta\right)d\Theta.
\end{equation}
Here, $\rho_{u}\left(U\right)$ and $\rho_{\phi}\left(\Theta\right)$
denote the system distribution functions in the representation
$R_{u}$ and $R_{\phi} $, respectively, which are mutually related by
the transformation rule:
\begin{equation}  \label{trans.ele.prob}
\rho_{\phi}\left(\Theta\right)=\rho_{u}\left(U\right)\left[\frac{%
\partial\Theta}{\partial U}\right]^{-1}.
\end{equation}

Let $dW$ be the elementary volume considering the number of
microstates belonging to the elementary subset
$\left[U,U+dU\right]$. As the case of the elementary probability
$dp$, $dW$ does not depend on the coordinate representation, and
hence, it obeys the following properties:
\begin{equation}
dW=\Omega_{u}\left(U\right)dU=\Omega_{\phi}\left(\Theta\right)d\Theta,
\end{equation}
\begin{equation}  \label{trans.omega}
\Omega_{\phi}\left(\Theta\right)=\Omega_{u}\left(U\right)\left[\frac{%
\partial\Theta}{\partial U}\right]^{-1}.
\end{equation}
Consequently, the probabilistic weight $\omega_{u}\left(U\right)$
considering the surroundings thermodynamic influence behaves as a
\textit{scalar function} under reparametrizations:
\begin{equation}  \label{trans.weight}
\omega_{u}\left(U\right)=\omega_{\phi}\left(\Theta\right)\equiv\omega_{u}
\left[U\left(\Theta\right)\right],
\end{equation}
with $U\left(\Theta\right)$ being the inverse function of
$\Theta\left(U\right)$.

The reparametrization invariance of the probability distribution
function also leads to the reparametrization invariance of the
expectation value of any physical observable $ A=A(U)=A(\Theta)$
(scalar function):
\begin{eqnarray}
\left\langle
A\right\rangle_{\phi}=\int_{\Theta_{inf}}^{\Theta_{sup}}A\left(\Theta\right)%
\rho_{\phi}\left(\Theta\right)d\Theta \\
=\int_{U_{inf}}^{U_{sup}}A\left(U\right)\rho_{\phi}\left(U\right)dU=\left%
\langle A\right\rangle_{u},
\end{eqnarray}
such that, one can denote the expectation values without indicating
the coordinate representation used for its expression:
\begin{equation}
\left\langle A\right\rangle_{u}=\left\langle
A\right\rangle_{\phi}\equiv\left\langle A\right\rangle.
\end{equation}

A remarkable equilibrium situation of the conventional
thermodynamics and statistical mechanics is the system in energetic
isolation, whose probabilistic weight:
\begin{equation}
\omega^{mic}_{u}\left(U|U_{0}\right)=\frac{1}{\Omega_{u}\left(U_{0}\right)}%
\delta\left(U-U_{0}\right)
\end{equation}
defines the known \textit{microcanonical ensemble}. This
probabilistic weight possesses the notable feature that its
mathematical form does not depend on the representation:
\begin{equation}
\omega^{mic}_{u}\left(U|U_{0}\right)=\omega^{mic}_{\phi}\left(\Theta|%
\Theta_{0}\right)=\frac{1}{\Omega_{\phi}\left(\Theta_{0}\right)}%
\delta\left(\Theta-\Theta_{0}\right),
\end{equation}
a property that is straightforwardly derived from the identity:
\begin{equation}
\delta\left(\Theta-\Theta_{0}\right)=\delta\left(U-U_{0}\right)\left[\frac{%
\partial \Theta\left(U_{0}\right)}{\partial U}\right]^{-1}
\end{equation}
and the transformation rule (\ref{trans.omega}).

\subsection{Reparametrization duality}

\label{dual}  Let us define the thermostat inverse temperature in
representation $R_{\phi}$ as:
\begin{equation}
\beta^{\phi}_{\omega}=-\frac{\partial\log\omega_{\phi}\left(\Theta\right)}{%
\partial \Theta}.
\end{equation}
Therefore, it obeys the transformation rule:
\begin{equation}
\beta^{\phi}_{\omega}=\beta^{u}_{\omega}\left[\frac{\partial \Theta}{%
\partial U}\right]^{-1}
\end{equation}
as a consequence of the scalar character of the probabilistic weight $\omega_{\phi}$, with $%
\beta^{u}_{\omega}$ being the thermostat (effective) inverse
temperature expressed in Eq.(\ref{beta.w}).

Boltzmann's entropy of the system in this representation can be
defined by:
\begin{equation}  \label{BE.g}
S_{\phi}=k_{B}\log W_{\phi},
\end{equation}
where $W_{\phi}=\Omega_{\phi}\delta\epsilon_{\phi}$, with $%
\delta\epsilon_{\phi}$ being a suitable constant that makes
$W_{\phi}$ dimensionless. The above coarsed-grained definition of
Boltzmann's entropy is
not properly a scalar function as the case of the probabilistic weight (\ref%
{trans.weight}). In fact, it obeys the transformation rule:
\begin{equation}  \label{trans.b.e}
S_{\phi}=S_{u}-k_{B}\log \left(\frac{\partial\Theta}{\partial U}\frac{%
\delta\epsilon_{u}}{\delta\epsilon_{\phi}}\right).
\end{equation}
As already pointed out by Ruppeiner (see subsection II.B of ref.\cite%
{rupper}), the density distribution function derived from Einstein's
postulate:
\begin{equation}  \label{Ein}
\rho_{x}(x)dx=C\exp\left[\frac{S\left(x\right)}{k_{B}}\right]dx
\end{equation}
obeys different mathematical forms under different coordinate
representations, $x\rightarrow y(x)$, if one assumes that the
entropy is a \textit{state function} whose value does not depend on
the representation
(scalar function), $S(x)=S(y)$. A simple analysis allows us to verify that the left-hand side of Eq.(\ref{Ein}%
) behaves as:
\begin{equation}
\rho_{x}(x)dx=\rho(x)\left|\frac{\partial x\left(y\right)}{\partial
y}\right|dy=\rho_{y}(y)dy,
\end{equation}
while its right-hand side as:
\begin{equation}
C\exp\left[\frac{S\left(x\right)}{k_{B}}\right]dx=C\exp\left[\frac{%
S\left(x\right)}{k_{B}}\right]\left|\frac{\partial
x\left(y\right)}{\partial y}\right|dy \neq
C'\exp\left[\frac{S\left(y\right)}{k_{B}}\right]dy.
\end{equation}
This fact not only constitutes an important defect in order to
develop a \textit{Riemannian formulation} of fluctuation theory, but
it also presupposes some inconsistences with the thermodynamic
arguments behind of Einstein's postulate for the fluctuation formula
of Eq.(\ref{Ein}). In this work, we shall assume the entropy
modification (\ref{trans.b.e}) associated with reparametrizations
and analyze its direct consequences. Clearly, such an alternative
definition allows us to preserve the functional dependence of
fluctuation formula (\ref{Ein}) in any coordinate representation. It
requires that the entropy is no longer a state function with a
scalar character, as is usually assumed in other geometric
formulations of fluctuation theory \cite{rupper}.

Under these above assumptions, the system inverse temperature
$\beta^{\phi}$ in the representation $R_{\phi}$ is given by:
\begin{equation}
\beta^{\phi}=\frac{\partial S_{\phi}}{\partial\Theta}
\end{equation}
and obeys the transformation rule:
\begin{equation}
\beta^{\phi}=\left[\beta^{u}-k_{B}\frac{\partial}{\partial U}\log \left(%
\frac{\partial\Theta}{\partial U}\frac{\delta\epsilon_{u}}{%
\delta\epsilon_{\phi}}\right)\right] \left[\frac{\partial \Theta}{\partial U}%
\right]^{-1}.
\end{equation}
As expected, the inverse temperature difference in the representation $%
R_{\phi}$ can be expressed as:
\begin{equation}  \label{inv.temp.dif.phi}
\eta_{\phi}\left(\Theta\right)=\beta^{\phi}_{\omega}\left(\Theta\right)-%
\beta^{\phi}\left(\Theta\right)=-k_{B}\frac{\partial\log\rho_{\phi}\left(%
\Theta\right)}{\partial \Theta}.
\end{equation}
By only admitting regular reparametrizations obeying the
constraints:
\begin{equation}
0<\left|\frac{\partial \Theta}{\partial U}\right|<+\infty,~0<\left|\frac{%
\partial^{2} \Theta}{\partial U^{2}}\right|<+\infty
\end{equation}
on every point $U\in M_{u}$, one can easily show the validity of the
boundary conditions:
\begin{equation}  \label{gen.b1}
\rho_{\phi}\left(\Theta_{inf}\right)=\rho_{\phi}\left(\Theta_{sup}\right)=0,
\end{equation}
\begin{equation}  \label{gen.b2}
\frac{\partial\rho_{\phi}\left(\Theta_{inf}\right)}{\partial \Theta}=\frac{%
\partial\rho_{\phi}\left(\Theta_{sup}\right)}{\partial \Theta}=0
\end{equation}
by starting from Eq.(\ref{bu1}) and Eq.(\ref{bu2}).

As already shown in the previous subsection, definition (\ref%
{inv.temp.dif.phi}) and the boundary conditions (\ref{gen.b1}) and (\ref%
{gen.b1}) lead to the following extensions of the rigorous identities (\ref%
{eq.cond}), (\ref{aux.fu}) and (\ref{gen.identity}):
\begin{equation}  \label{eq.cond.2}
\left\langle\eta_{\phi}\right\rangle=\int^{\Theta_{sup}}_{\Theta_{inf}}%
\eta_{\phi}\left(\Theta\right)\rho_{\phi}\left(\Theta\right)d\Theta=0,
\end{equation}
\begin{equation}  \label{aux.fu.2}
\left\langle
\Theta\eta_{\phi}\right\rangle=\int^{\Theta_{sup}}_{\Theta_{inf}}\Theta%
\eta_{\phi}\left(\Theta\right)\rho_{\phi}\left(\Theta\right)d\Theta=k_{B},
\end{equation}
\begin{equation}  \label{comple.flu.3}
\left\langle -k_{B}\frac{\partial A}{\partial \Theta}+A\eta_{\phi}\right%
\rangle=0,
\end{equation}
as well as the generalized fluctuation theorems:
\begin{equation}  \label{mother.2}
\left\langle \delta\Theta\delta\eta_{\phi} \right\rangle=k_{B},
\end{equation}
\begin{equation}  \label{comple.flu.2}
\left\langle -k_{B}\frac{\partial\eta_{\phi}}{\partial \Theta}%
+\delta\eta_{\phi}^{2}\right\rangle=0,
\end{equation}
and finally, the thermodynamic uncertainty relation:
\begin{equation}  \label{unc.2}
\Delta\Theta\Delta\eta_{\phi} \geq k_{B}.
\end{equation}

Thus, the consideration of coordinate changes makes it possible to
extend the results already derived by using the energy
representation
$R_{u}$. Although the thermodynamic identities (\ref{eq.cond},\ref{aux.fu}) and (%
\ref{eq.cond.2},\ref{aux.fu.2}), and the fluctuation theorems (\ref{mother},\ref%
{comple.flu}) and (\ref{mother.2},\ref{comple.flu.2}), as well as
the uncertainty relations (\ref{unc}) and (\ref{unc.2}) are closely
related, \textit{they represent different thermodynamic relations
characterizing the same equilibrium situation}. It could be said
that all of these mutually related identities account for the
existence of  a special kind of internal symmetry, which shall be
hereafter referred to as \textit{reparametrization duality}.

The invariance under reparametrizations (coordinate transformation
or \textit{diffeomorphisms}) is the same kind of symmetry considered
by Einstein's theory of gravitation. However, there exist radical
differences between this latter physical theory and the geometric
statistical formalism developed in this work. (1) While the
gravitation theory is defined in terms of \textit{local quantities},
the rigorous thermodynamic identities obtained here are expressed in
terms of \textit{statistical expectation values} defined over the
entire subset $M_{\phi}$ representing all admissible system
macrostates in the present equilibrium situation, that is, this is a \textit{%
non-local theory}, similar to quantum mechanics. (2) Furthermore,
Einstein's theory refers to the same physical laws in different
representations, while the above thermodynamic identities consider a
family of \textit{different fluctuations relations} exhibiting the
same mathematical appearance under different coordinate
representations of a given equilibrium situation. This is why we
refer to it as reparametrization duality instead of
reparametrization symmetry.

In the next subsection, we shall arrive at a local formulation of
the present approach with a Riemannian-like structure closely
related to other
geometric approaches of fluctuation theory existing in the literature \cite%
{rupper}. We shall see, however, that such a development presupposes
the consideration of certain unexpected approximations.

\subsection{Riemannian approach}

\label{Gaussian.sub} Let us assume that the systems under
consideration are large enough to deal with the thermodynamic
fluctuations by using a \textit{Gaussian approximation}. An
essential assumption considered here is that the system undergoes
small thermal fluctuations close to its
equilibrium point, which is determined by \textit{the most likely macrostate}%
.

A problem encountered is that \textit{the most likely macrostate
actually depends on the coordinate representation used for
describing the system behavior}, which is a direct consequence of
the non-scalar character of the system entropy. In order to show
this fact, let us consider the transformation rule of the inverse
temperature difference:
\begin{equation}  \label{trans.eta}
\eta_{\phi}=\frac{\partial U}{\partial \Theta}\left[\eta_{u}+k_{B}\frac{%
\partial}{\partial U}\log \left(\frac{\partial\Theta}{\partial U}\frac{%
\delta\epsilon_{u}}{\delta\epsilon_{\phi}}\right)\right].
\end{equation}
The stationary condition associated with the most likely macrostate
in each representation are given by:
\begin{equation}
\eta_{u}\left(\bar{U}\right)=0~for~R_{u}~and~\eta_{\phi}\left(%
\bar{\Theta}\right)=0~for~R_{\phi},
\end{equation}
According to Eq.(\ref{trans.eta}), the vanishing of $\eta_{u}$ does
not correspond to a vanishing of $\eta_{\phi}$, and vice versa, a
result showing that the most likely macrostate depends on the
coordinate representation.

This last result contracts with the general validity of the thermal
equilibrium condition in terms of statistical expectation values,
Eq.(\ref {eq.cond.2}). It clearly indicates that the method
generally used for deriving such a condition in terms of the most
likely macrostate is just a suitable approximation. Nevertheless, it
could be easily noticed that the modification involved during the
reparametrization change is just a second-order effect. The
transformation rule (\ref{trans.eta}) can be combined with
Eq.(\ref{eq.cond}) and Eq.(\ref{eq.cond.2}) in order to obtain:
\begin{equation}  \label{res.likely}
\left\langle \delta\eta_{\phi}\delta \Lambda^{\phi}_{u}
\right\rangle=\left\langle k_{B}\frac{\partial}{\partial U}\log
\left(\Lambda^{\phi}_{u}\frac{\delta\epsilon_{u}}{\delta\epsilon_{\phi}}%
\right)\right\rangle,
\end{equation}
where the following notation is considered:
\begin{equation}
\Lambda^{\phi}_{u}=\frac{\partial\Theta\left( U\right)}{\partial U}.
\end{equation}
Eq.(\ref{res.likely}) indicates that the second additive term on
right-hand side of the transformation rule (\ref{trans.eta}) is just
a small correction, which can be disregarded in most practical
applications. Therefore, one can admit the approximate relation:
\begin{equation}  \label{eq.app}
\bar{\eta}_{\phi}=\bar{\eta}_{u}\left(\bar{\Lambda}^{\phi}_{u}\right)^{-1}%
\equiv0,
\end{equation}
where $\bar{A}$ denotes the value of the function $A\left(U\right)$
at the most likely macrostate, $\bar{A}\equiv A\left(\bar{U}\right)$.

Basically, the approximation assumed in Eq.(\ref{eq.app}) is
equivalent to considering Boltzmann's entropy (\ref{BE.g}) as a
scalar function, and hence, the approximate transformation rule of
the system inverse temperature is given by:
\begin{equation}
\bar{\beta}^{\phi} = \bar{\beta}^{u} \left( \bar{\Lambda}^{\phi}_{u}
\right)^{-1} .
\end{equation}

In general, the Gaussian approximation allows us to consider the
fluctuations of an arbitrary energy function $A\left(U\right)$ as:
\begin{equation}  \label{gaussian}
\delta A=\frac{\partial A\left(\bar{U}\right)}{\partial U}\delta U.
\end{equation}
In particular, it allows us to introduce the following
transformation rule:
\begin{equation}  \label{trans.dfi}
\delta\Theta=\bar{\Lambda}^{\phi}_{u}\delta U .
\end{equation}
Moreover, by starting from Eq.(\ref{eq.app}), we obtain:
\begin{equation}
\delta\eta_{\phi}=\left(\bar{\Lambda}^{\phi}_{u}\right)^{-1}\left[%
\delta\eta_{u}-\bar{\eta}_{u}\frac{\partial}{\partial U}\log\left(\bar{%
\Lambda}^{\phi}_{u}\frac{\delta\epsilon_{u}}{\delta\epsilon_{\phi}}%
\right)\delta U\right],
\end{equation}
which reduces to:
\begin{equation}
\delta\eta_{\phi}=\left(\bar{\Lambda}^{\phi}_{u}\right)^{-1}\delta\eta_{u},
\end{equation}
after considering the thermal equilibrium condition
$\bar{\eta}^{u}=0$. Using this latter transformation rules, one can
obtain the transformation rules of some fluctuations relations:
\begin{equation}  \label{cont.tens}
\left\langle \delta\Theta^{2}\right\rangle=\left(\bar{\Lambda}%
^{\phi}_{u}\right)^{2}\left\langle \delta U^{2}\right\rangle
\end{equation}
\begin{equation}  \label{cov.tens}
\left\langle \delta\eta_{\phi}^{2}\right\rangle=\left(\bar{\Lambda}%
^{\phi}_{u}\right)^{-2}\left\langle \delta \eta_{u}^{2}\right\rangle
\end{equation}
\begin{equation}  \label{scal.tens}
\left\langle \delta
\Theta\delta\eta_{\phi}\right\rangle=\left\langle \delta U\delta
\eta_{u} \right\rangle\equiv k_{B}.
\end{equation}

Exactly, Eqs.(\ref{cont.tens})-(\ref{scal.tens}) correspond to
transformation rules of contravariant second-rank tensors, covariant
second-range tensor and scalar functions in a differential geometric
theory, respectively. In order to provide a \textit{Riemannian
structure} to the present geometrical approach, we must introduce an
appropriate \textit{metric}. Such a role could be carried out by the
\textit{global curvature} $K_{\phi}$:
\begin{equation}
K_{\phi}=\frac{\partial\eta_{\phi}}{\partial\Theta}=-k_{B}\frac{%
\partial^{2}\log\rho_{\phi}}{\partial \Theta^{2}}
\end{equation}
evaluated at the most likely macrostate, which allows for the
conversion between the fluctuations of the conjugated thermodynamic
quantities (covariant and contravariant vectors) within the Gaussian
approximation:
\begin{equation}
\delta\eta_{\phi}=\bar{K}_{\phi}\delta\Theta.
\end{equation}
The global curvature obeys the transformation rule:
\begin{equation}  \label{gen.glob.curv.trans}
K_{\phi}=\left(\Lambda^{\phi}_{u}\right)^{-2}\left\{K_{u}-\eta_{u}\frac{%
\partial c_{\phi}}{\partial U}+k_{B}\left[\frac{\partial^{2} c_{\phi}}{%
\partial U^{2}}-\left(\frac{\partial c_{\phi}}{\partial U}\right)^{2}\right]%
\right\}
\end{equation}
with $c_{\phi}=\log\left(\Lambda^{\phi}_{u}\delta\epsilon_{u}/\delta%
\epsilon_{\phi}\right)$, which reduces to:
\begin{equation}
\bar{K}_{\phi}=\left(\bar{T}^{\phi}_{u}\right)^{-2}\bar{K}_{u}
\end{equation}
after considering the thermal equilibrium condition
$\bar{\eta}_{u}=0$ and dismissing small contributions associated
with the non-scalar character of Boltzmann's entropy (the two terms
associated with the Boltzmann's constant $k_{B}$). Clearly, the
global curvature can only be considered as a second-rank covariant
tensor under the above approximations, since the general
transformation rule (\ref{gen.glob.curv.trans}) does not correspond
to this kind of geometric object. Interestingly, such a function
appears in the complementary fluctuation relation
(\ref{comple.flu.2}), which establishes the non-negative character
of its expectation value in any coordinate representation:
\begin{equation}
k_{B}\left\langle K_{\phi}\right\rangle= \left\langle
\delta\eta^{2}_{\phi} \right\rangle .
\end{equation}
As already commented, this rigorous fluctuation relation satisfies,
as a whole, the reparametrization duality, which is not the case of
the global curvature $K_{\phi}$ considered as an individual entity.

By using the global curvature $\bar{K}_{\phi}$, one can easily
obtain other fluctuations relations such as:
\begin{equation}
\left\langle\delta\eta^{\phi}_{\omega}\delta\Theta\right\rangle=\bar{K}%
_{\phi}\left\langle\delta\Theta^{2}\right\rangle=\bar{K}_{u}\left\langle%
\delta U^{2}\right\rangle=\left\langle\delta\eta^{u}_{\omega}\delta
U\right\rangle=k_{B},
\end{equation}
and rewrite the distribution function $\rho_{\phi}$ in this Gaussian
approximation as follows:
\begin{equation}
\rho_{\phi}\left(\Theta|\bar{\Theta}\right)d\Theta=\sqrt{\frac{\bar{K}_{\phi}%
}{2\pi k_{B}}}\exp\left[-\frac{1}{2k_{B}}\bar{K}_{\phi}\left(\Theta-\bar{%
\Theta}\right)^{2}\right]d\Theta .
\end{equation}

\section{Generalized Gibbs canonical ensemble}

Let us denote by $T_{\phi}$ the thermostat temperature in the
representation $R_{\phi}$, with $\beta^{\phi}_{\omega}=1/T_{\phi}$.
One can formally introduce the heat capacity $C_{\phi}$ of this
representation as:
\begin{equation}
C_{\phi}=\frac{\partial \Theta}{\partial T_{\phi}},
\end{equation}
which allows us to obtain a geometric extension of
fluctuation-dissipation relation (\ref{fdr}):
\begin{equation}  \label{fdr2}
k_{B}\bar{C}_{\phi}=\left(\bar{\beta}^{\phi}_{\omega}\right)^{2}\left\langle%
\delta\Theta^{2}\right\rangle+\bar{C}_{\phi}\left\langle\delta\beta^{\phi}_{%
\omega}\delta\Theta\right\rangle
\end{equation}
after combining the Gaussian approximation:
\begin{equation}
\delta\beta_{\phi}=-\bar{\beta}^{2}_{\phi}/\bar{C}_{\phi}\delta\Theta
\end{equation}
with definition (\ref{inv.temp.dif.phi}) and the fluctuation relation (\ref%
{mother.2}). A relevant case among the admissible equilibrium
situations considered by the above fluctuation-dissipation relation
is the one obeying the constraint
$\delta\beta^{\phi}_{\omega}\equiv0$, which is associated with the
following distribution function:
\begin{equation}
dp_{c}\left(\Theta|\beta^{\phi}_{c}\right)=\frac{1}{Z\left(\beta^{\phi}_{c}%
\right)}\exp\left(-\frac{1}{k_{B}}\beta^{\phi}_{c}\Theta\right)\Omega_{\phi}%
\left(\Theta\right)d\Theta.
\end{equation}
This is just the analogous version of the Gibbs canonical ensemble in the $%
R_{\phi}$ representation, with $\beta^{\phi}_{c}$ being a constant
parameter. By rewriting this particular distribution function in the
energy representation $R_{u}$:
\begin{equation}  \label{gen.can}
dp_{c}\left(U|\beta^{\phi}_{c}\right)=\frac{1}{Z\left(\beta^{\phi}_{c}\right)%
}\exp\left[-\frac{1}{k_{B}}\beta^{\phi}_{c}\Theta\left(U\right)\right]%
\Omega_{u}\left(U\right)dU
\end{equation}
one arrives at the same expression found for the so-called
\textit{generalized canonical ensemble} recently proposed in the
literature \cite{hugo.gce,toral}. Let us now analyze its general
mathematical properties.

\subsection{General mathematical properties}

As usual, the partition function $Z\left(\beta^{\phi}_{c}\right)$
derived from the normalization condition:
\begin{equation}  \label{planck}
Z\left(\beta^{\phi}_{c}\right)=\int_{U_{inf}}^{U_{sup}} e^{-\frac{1}{k_{B}}%
\beta_{c}^{\phi}\Theta\left(U\right)}\Omega_{u}\left(U\right)dU
\end{equation}
allows us to obtain the \textit{generalized Planck's thermodynamic
potential}:
\begin{equation}
P_{\phi}\left(\beta^{\phi}_{c}\right)=-k_{B}\log
Z\left(\beta^{\phi}_{c}\right),
\end{equation}
which provides two relevant statistical expectation values:
\begin{equation}
\left\langle \Theta\right\rangle=\frac{\partial
P_{\phi}\left(\beta_{c}^{\phi}\right)}{\partial\beta_{c}^{\phi}}%
,~\left\langle
\delta\Theta^{2}\right\rangle=-k_{B}\frac{\partial^{2}
P_{\phi}\left(\beta_{c}^{\phi}\right)}{\partial(\beta_{c}^{\phi})^{2}}.
\end{equation}
These last results can be combined in order to obtain the canonical
version of the fluctuation-dissipation relation (\ref{fdr2}):
\begin{equation}
-k_{B}\frac{\partial\left\langle \Theta\right\rangle}{\partial\beta_{c}^{\phi}}%
=\left\langle \delta\Theta^{2}\right\rangle\Rightarrow
k_{B}C^{c}_{\phi}=(\beta_{c}^{\phi})^{2}\left\langle
\delta\Theta^{2}\right\rangle,
\end{equation}
with $C^{c}_{\phi}$ being the canonical heat capacity:
\begin{equation}
C^{c}_{\phi}=\frac{\partial \left\langle
\Theta\right\rangle}{\partial T_{\phi}}.
\end{equation}
Clearly, this theorem states that the stable thermodynamically
macrostates are those with a nonnegative heat capacity
$C^{c}_{\phi}>0$.

Let us now rewrite Planck's thermodynamic potential in the
$R_{\phi}$ representation:
\begin{eqnarray}
e^{-P_{\phi}\left(\beta^{\phi}_{c}\right)/k_{B}}=\int_{\Theta_{inf}}^{%
\Theta_{sup}}
e^{-\frac{1}{k_{B}}\beta_{c}^{\phi}\Theta}\Omega_{\phi}\left(\Theta\right)d\Theta \\
=\int_{\Theta_{inf}}^{\Theta_{sup}}
e^{-\frac{1}{k_{B}}\left[\beta_{c}^{\phi}\Theta-S_{\phi}\left(\Theta\right)\right]}\frac{d\Theta}{%
\delta\epsilon_{\phi}}
\end{eqnarray}
and develop a Gaussian approximation (the second-order power expansion in $%
\Theta$) around the local maxima:
\begin{equation}
\simeq e^{-P_{\phi}^{*}/k_{B}}\int_{\Theta_{inf}}^{\Theta_{sup}} e^{-\frac{1%
}{2k_{B}}\kappa^{*}_{\phi}\Delta\Theta^{2}}\frac{d\Theta}{%
\delta\epsilon_{\phi}}
\end{equation}
with $\Delta\Theta=\Theta-\Theta_{c}$ and $P^{*}_{\phi}$ given by:
\begin{equation}  \label{legendre}
P^{*}_{\phi}=\inf_{\Theta_{s}}\left\{\beta_{c}^{\phi}\Theta-S_{\phi}\left(%
\Theta\right)\right\}.
\end{equation}
The local maxima $\Theta_{c}$ are derived from the stationary
and stability conditions:
\begin{equation}
\beta_{c}^{\phi}=\frac{\partial
S_{\phi}\left(\Theta_{s}\right)}{\partial
\Theta}\equiv\beta^{\phi}\left(\Theta_{c}\right),~\kappa^{*}_{\phi}=-\frac{%
\partial^{2} S_{\phi}\left(\Theta_{s}\right)}{\partial \Theta^{2}}>0.
\end{equation}
By admitting the existence of only one maximum, this approximation
yields:
\begin{eqnarray}
P_{\phi}\left(\beta^{\phi}_{c}\right)\simeq P^{*}_{\phi}-\frac{1}{2}%
\log\left(\frac{2\pi k_{B}}{\kappa^{*}_{\phi}\delta\epsilon^{2}_{\phi}}%
\right), \\
\left\langle\Delta\Theta^{2}\right\rangle=k_{B}\frac{1}{\kappa^{*}_{\phi}}.
\end{eqnarray}
Clearly, the additive logarithmic term in the Gaussian estimation of
the Planck thermodynamic potential constitutes a small correction in
the case of sufficiently large systems. By dismissing this small
contribution, one finds that Planck's thermodynamic potential is approximately given by the known \textit{%
Legendre transformation}:
\begin{equation}  \label{legendre}
P^{*}_{\phi}\left(\beta^{\phi}_{c}\right)=\inf_{\Theta_{s}}\left\{%
\beta_{c}^{\phi}\Theta-S_{\phi}\left(\Theta\right)\right\}.
\end{equation}
The stationary condition is merely the condition of thermal
equilibrium associated with this representation, while the stability
condition is simply the
requirement of non-negativity of the microcanonical heat capacity $C_{\phi}$%
:
\begin{equation}  \label{c.may.z}
\frac{\partial^{2} S_{\phi}\left(\Theta_{s}\right)}{\partial \Theta^{2}}%
=-\left(\beta^{\phi}\right)^{2}\frac{1}{C_{\phi}}<0\Rightarrow
C_{\phi}>0.
\end{equation}

Eqs.(\ref{planck})-(\ref{c.may.z}) correspond to many well-known
dual expressions previously obtained within the Gibbs canonical
ensemble (\ref{gibbs}). Obviously, these two ensembles are
intimately related. By considering the scalar character of the
probabilistic weight $\omega_{\phi}$:
\begin{equation}
\omega_{\phi}\left(\Theta|\beta^{u}_{c}\right)=\frac{1}{Z\left(\beta^{%
\phi}_{c}\right)}\exp\left(-\frac{1}{k_{B}}\beta^{\phi}_{c}\Theta\right),
\end{equation}
the thermostat inverse temperature $\beta^{u}_{\omega}$ in the energy representation $R_{u}$ is given by:
\begin{equation}
\beta^{u}_{\omega}\left(U\right)=\beta^{\phi}_{c}\frac{\partial
\Theta\left(U\right)}{\partial U}.
\end{equation}
This latter result clarifies that the generalized canonical ensemble
(\ref{gen.can}) corresponds to a special kind of equilibrium
situation with a variable (fluctuating) inverse temperature of all
admissible states accounted for by fluctuation-dissipation relation
(\ref{fdr}), that is, a situation with
non-vanishing system-surroundings correlative effects $\left\langle\delta%
\beta^{u}_{\omega}\delta U\right\rangle\neq0$.

By considering the transformation rule for the microcanonical curvature $%
\kappa_{\phi}$:
\begin{equation}  \label{trans.kap}
\kappa_{\phi}=\left(\Lambda^{\phi}_{u}\right)^{-2}\left\{\kappa_{u}+\beta^{u}%
\frac{\partial c_{\phi}}{\partial U}+k_{B}\left[\frac{\partial^{2} c_{\phi}}{%
\partial U^{2}}-\left(\frac{\partial c_{\phi}}{\partial U}\right)^{2}\right]%
\right\},
\end{equation}
one can find that the requirement $\kappa_{\phi}>0$ can be combined
with the existence of macrostates with $\kappa_{u}<0$ in the energy
representation with an appropriate selection of the reparametrization $%
U\rightarrow\Theta\left(U\right)$\footnote{%
The presence of additive terms with Boltzmann's factor $k_{B}$ in Eq.(\ref%
{trans.kap}) takes into account the modification of the system
entropy during a reparametrization and the consequent correction of
the most likely macrostate.} . This fact is more evident when
working in the energy representation $R_{u}$, where the stability
condition reads as follows:
\begin{equation}
\bar{\kappa}_{u}+\beta^{u}_{c}\frac{\partial^{2} \Theta\left(\bar{U}\right)}{%
\partial U^{2}}=\bar{\kappa}_{u}+\frac{\partial \beta^{u}_{\omega}\left(\bar{%
U}\right)}{\partial U}>0.
\end{equation}
By considering the relations $\kappa_{u}=(\beta^{u})^{2}/C_{u}$ and
$\partial \beta^{u}_{\omega}/\partial U
=(\beta^{u}_{\omega})^{2}/C^{u}_{\omega}$, with $C_{u}$ and
$C^{u}_{\omega}$ being the heat capacities of the system and the
thermostat respectively (their usual definitions), as
well as by using the thermal equilibrium condition $\bar{\beta}^{u}=\bar{%
\beta}^{u}_{\omega}=\beta$, one arrives at the expression:
\begin{equation}
\frac{C_{u}C^{u}_{\omega}}{C^{u}_{\omega}+C_{u}}>0,
\end{equation}
which leads to Thirring's stability condition (\ref%
{c.thir}) for macrostates with $C_{u}<0$.

As the Gibbs canonical ensemble (\ref{gibbs}), the present geometric
extension (\ref{gen.can}) becomes equivalent to the microcanonical
ensemble with increasing of the system size $N$,
$\Delta\Theta/\Theta\sim 1/\sqrt{N}$, an equivalency that can be
ensured even for macrostates with $C_{u}<0$ or $\kappa_{u}<0$ with
an appropriate selection of the reparametrization
$\Theta\left(U\right)$. This remarkable property makes this ensemble
a very attractive thermo-statistical framework, since besides of
exhibiting many notable properties of the usual the Gibbs canonical
ensemble, it also provides a better treatment of the phenomenon of
ensemble inequivalence associated with the presence of negative heat
capacities, as already discussed in refs.\cite{hugo.gce,toral}. In
particular, this statistical ensemble constitutes a suitable
framework for extending of Monte Carlo methods, as discussed in
subsection \ref{Monte}.

\subsection{Derivation from information theory}

It is possible to realize that the generalized Gibbs canonical ensemble (%
\ref{gen.can}) can also be derived from Jaynes's reinterpretation of
statistical mechanics in terms of the information theory of Shannon
\cite {jaynes}, e.g., by considering the maximization of the known
statistical (extensive) information entropy:
\begin{equation}  \label{ent}
S_{e}=-\sum_{k} p_{k} \log p_{k},
\end{equation}
under the normalization condition:
\begin{equation}  \label{const.fi}
\left\langle 1\right\rangle=\sum_{k}
p_{k}=1
\end{equation}
and the following nonlinear energy-like constraint:
\begin{equation}  \label{const.fi}
\left\langle \Theta\right\rangle=\sum_{k}
\Theta\left(U_{k}\right)p_{k}.
\end{equation}
Such a derivation was developed by Toral in ref.\cite{toral}. The
interested reader can refer to this work for more details.

Clearly, the bijective character of the reparametrization $%
U\leftrightarrow\Theta\left(U\right)$ should ensure that this
generalized ensemble exhibits almost the same stationary properties
obtained from the application of the Gibbs canonical ensemble in
sufficiently large systems, where one usually assumes the
appropriateness of the Gaussian approximation. However, the
nonlinear character of the bijective application
$\Theta\left(U\right)$ produces a deformation in the canonical
description, which conveniently modifies the system fluctuating
behavior and the accessible regions of the subset of all admissible
system macrostates $M_{u}$.

\subsection{Connections with inference theory: generalization of
Mandelbrot's approach}

Generally speaking, \textit{statistical inference} can be described
as the
problem of deciding how well a set of outcomes $\left(x_{1},x_{2},%
\ldots,x_{m},\right)$, obtained from independent measurements, fits to
a proposed probability distribution:
\begin{equation}
dp\left(x|\theta\right)=\rho\left(x|\theta\right)dx.
\end{equation}
If the probability distribution is characterized by one or more parameters ($%
\theta$), this problem is equivalent to inferring the value of the
parameter(s) from the observed measurement outcomes $x$. To make
inferences about the parameter, one constructs estimators, i.e.,
functions:
\begin{equation}
\hat{\theta}\left(x_{1},x_{2},\ldots,x_{m}\right)
\end{equation}
of the outcomes of $m$ independent repeated measurements
\cite{Fisher}. The value of this function represents the best guess
for $\theta$.

Commonly, there exist several criteria imposed on estimators in
order to ensure that their values constitute \textit{good} estimates
of the parameter $\theta$, such as:

\begin{itemize}
\item \textit{Unbiasedness}:
\begin{equation}\label{Unbiasedness}
\left\langle\hat{\theta}\right\rangle=\int \hat{\theta}\left(x_{1},x_{2},%
\ldots,x_{m}\right)\prod^{m}_{k=1}dp\left(x_{k}|\theta\right)=\theta.
\end{equation}

\item \textit{Efficiency} or minimal statistical dispersion:
\begin{equation}\label{Efficiency}
\left\langle\delta\hat{\theta}^{2}\right\rangle=\int \left(\hat{\theta}%
-\left\langle\hat{\theta}\right\rangle\right)^{2}\prod^{m}_{k=1}dp%
\left(x_{k}|\theta\right)\rightarrow minimum.
\end{equation}

\item \textit{Sufficiency}:
\begin{equation}\label{Sufficiency}
dp\left(x_{1},x_{2},\ldots,x_{m}|\theta\right)=f\left(x_{1},x_{2},%
\ldots,x_{m}\right)dp\left(\hat{\theta}\right),
\end{equation}
where $dp\left(\hat{\theta}\right)$ is the marginal distribution of $\hat{%
\theta}$ and $f\left(x_{1},x_{2},\ldots,x_{m}\right)$ is an
arbitrary function of the measurements, independent on $\theta$.
\end{itemize}

Since any statistical estimator $\hat{\theta}$ represents a
stochastic quantity, it is natural in inference problems that an
estimator obeys the unbiasedness (\ref{Unbiasedness}) and efficiency
(\ref{Efficiency}) conditions. However, there exists a remarkable
theorem of inference theory, the \textit{Cram\'{e}r-Rao's inequality}%
, which places an inferior bound on the efficiency of an arbitrary
unbiased estimator:
\begin{equation}  \label{Cramer.Rao}
\left\langle\delta\hat{\theta}^{2}\right\rangle\geq\frac{1}{%
I_{F}\left(\theta\right)},
\end{equation}
where $I_{F}\left(\theta\right)$ is the so-called \textit{Fisher's
information entropy}:
\begin{equation}  \label{Fisher}
I_{F}\left(\theta\right)=\int \left[\frac{\partial
\log\rho\left(x|\theta\right)}{\partial\theta}\right]^{2}\rho\left(x|\theta%
\right)dx.
\end{equation}

On the other hand, efficiency condition (\ref{Sufficiency}) ensures that, given the value of $%
\hat{\theta}\left(x_{1},x_{2},\ldots,x_{m}\right)$, the values of the data $%
\left(x_{1},x_{2},\ldots,x_{m}\right)$ are distributed independently of $%
\theta$, containing in this way all of the information about parameter $%
\theta$ that can be obtained from the data. As with unbiasedness and
efficiency, sufficiency is also a natural desirable condition in
inference problems. However, a theorem by Pitman and Koopman
\cite{Koopman} states that sufficient estimators only exist for a
reduced family of distribution functions, the so-called
\textit{exponential family}:
\begin{equation}  \label{PK}
dp\left(x|\theta\right)=\exp\left[A\left(\theta\right)+B\left(x\right)C%
\left(\theta\right)+D\left(x\right)\right]dx.
\end{equation}

Mandelbrot was the first investigator to realize the intimate
connection between statistical mechanics and inference theory
\cite{mandelbrot.it}. Clearly, the Gibbs canonical ensemble
(\ref{gibbs}) constitutes a relevant physical example of
probabilistic distribution function belonging to the exponential
family (\ref{PK}). As the well-known Kinchin work in the framework
of information theory \cite{Kinchin}, Mandelbrot proposed a set of
axioms in order to justify a direct derivation of the Gibbs
canonical ensemble in the framework of inference theory. Moreover,
he also focussed the inference problem of the inverse temperature
$\beta$, which appears as a parameter of the Gibbs canonical
ensemble (\ref{gibbs}), through some an unbiased estimator
$\hat{\beta} $ defined for a set of outcomes of the system energy
$U$. Thus, this author provided an interpretation of the
\textit{energy-temperature complementarity} previously postulated by
Bohr and Heisenberg \cite{bohr,Heisenberg}:
\begin{equation}  \label{Mandel.TUR}
\Delta_{c}\hat{\beta}\Delta_{c} U\geq k_{B},
\end{equation}
with $\Delta_{c} x\equiv\sqrt{\left\langle\delta
x^{2}\right\rangle}_{c}$, a result that follows from the
Cram\'{e}r-Rao's inequality (\ref{Cramer.Rao})
after noting that the Fisher's information entropy (\ref%
{Fisher}) for the Gibbs canonical distribution (\ref{gibbs}) is
simply the canonical expectation value $\left\langle
\ast\right\rangle_{c}$ of the energy dispersion,
$I_{F}\left(\beta\right)\equiv\left\langle\delta
U^{2}\right\rangle_{c}$.

After reading the present discussion, one can point out some
critiques to Mandelbrot's approach. In regard to his interpretation
of energy-temperature complementarity, Eq.(\ref{Mandel.TUR}), it is
clear that such an uncertainty relation only applies in the
framework of the Gibbs canonical ensemble (\ref{gibbs}). Moreover,
this inequality accounts for the
limits of precision of a statistical estimation of the inverse temperature $%
\beta$ appearing as a parameter of the canonical ensemble
(\ref{gibbs}). Clearly, this quantity has nothing to do with the
system inverse temperature, but rather the inverse temperature of
the Gibbs thermostat. This is a common misunderstanding of some
contemporary developments of statistical physics, where it is not
distinguished between these two temperatures, leading in this way to
some limitations and inconsistences. Clearly, such difficulties are
overcome by the uncertainty relation (\ref{unc}) associated with the
energy-temperature fluctuation-dissipation relation (\ref{fdr}).

The differences between the Gibbs temperature of the canonical
ensemble (\ref{gibbs}) and the Boltzmann's definition (\ref{bb}) are
irrelevant in the case of large short-range thermodynamic systems
considered in conventional applications of statistical mechanics and
thermodynamics, overall, in those physical situations where the
necessary conditions for the equivalence between canonical and
microcanonical descriptions apply. However, the existing differences
become critical when one considers the thermodynamical description
of long-range interacting systems such as the astrophysical ones,
where the presence of macrostates with \textit{negative heat
capacities} constitutes an important thermodynamic feature that
rules their macroscopic behavior and dynamical evolution
\cite{gro1,Dauxois}. As already discussed, such an anomaly
cannot be described by using the Gibbs canonical description (\ref{gibbs}%
). Besides, there does not exist in this context an appropriate
Gibbs thermostat that ensures the existence of a thermal contact (a
boundary interaction) in presence of a long-range interacting force
such as gravity.

The above limitations also extend to other physical contexts such as
small or mesoscopic nuclear, molecular and atomic clusters, where
the presence of a negative heat capacity is not an unusual feature
\cite{gro1}, while the thermodynamic influence of a Gibbs thermostat
constitutes a very strong perturbation of its internal thermodynamic
state. In this kind of scenario, there does not always exist a clear
justification for the direct application of some theoretical
developments based on the consideration of the Gibbs canonical
ensemble, e.g.: the use of finite-temperature calculations for the
study of collisions in high energy physics. Interestingly, a
collective phenomenon such as the \textit{nuclear
multi-fragmentation} resulting from collisions of heavy nuclei is
simply a first-order phase transition revealing the experimental
observation of macrostates with negative heat capacities $C<0$
\cite{moretto,Dagostino}. Clearly, such a realistic phenomenon
cannot be appropriately described by using the canonical ensemble.

Remarkably, its is easy to note that the Gibbs canonical ensemble (%
\ref{gibbs}) is not the only one probabilistic distribution function
justified in terms of inference theory, as originally presupposed by
Mandelbrot in his approach. In fact, the whole family of the
generalized Gibbs canonical ensembles (\ref{gen.can}) also belongs
to the exponential family (\ref{PK}), and hence, such distributions
also ensure the existence of sufficient estimators
$\hat{\beta}^{\phi}_{c}$ obeying uncertainty relations \textit{\'{a}
la Mandelbrot}:
\begin{equation}  \label{Mandel.tur.fi}
\Delta_{\phi}\hat{\beta}^{\phi}_{c}\Delta_{\phi}\Theta\geq k_{B},
\end{equation}
as a consequence of the underlying reparametrization duality
discussed in this work. As expected, $\Delta_{\phi}x\equiv
\sqrt{\left\langle\delta x^{2}\right\rangle_{\phi}}$, with
$\left\langle \ast\right\rangle_{\phi}$ being the generalized
canonical expectation values derived from the generalized ensemble
(\ref{gen.can}).

\section{Conclusions}
We have provided in this work a panoramic overview of direct
implications and connections of the energy-temperature
fluctuation-dissipation relation (\ref{fdr}) with different
challenging questions of statistical mechanics.

As briefly discussed, the main motivation and most direct
consequence of this generalized fluctuation relation was the
compatibility with macrostates having negative heat capacities in
the framework of fluctuation theory. Such a feature makes possible
to analyze and apply the necessary conditions for the
thermodynamical stability of such anomalous macrostates in order to
extend the available Monte Carlo methods based on the consideration
of the Gibbs canonical ensemble (\ref{gibbs}), a procedure that also
allows one to avoid the incidence of the so-called
\textit{super-critical slowing down} encountered in large-scale
simulations. Moreover, the fluctuation-dissipation relation
constitutes a particular expression of a fluctuation relation
leading to the existence of a complementary relationship between
thermodynamic quantities of energy and (inverse) temperature
(\ref{unc}).

The consideration of geometric concepts, such as coordinate changes
or \textit{reparametrizations}, leads to a direct extension of many
old and new rigorous results of statistical mechanics in terms of a
special kind of internal symmetry that we refer to here as a
\textit{reparametrization duality}. Such a basis inspires the
introduction of a geometric generalized version of the Gibbs
canonical ensemble (\ref{gen.can}), which has been recently proposed
in the literature \cite{hugo.gce,toral}. This latter probabilistic
distribution allows for a better treatment of the phenomenon of
ensemble inequivalence or for the consideration of anomalous
macrostates with negative heat capacities. At the same time, this
family of distribution functions still preserves many notable
properties of the Gibbs canonical ensemble, including its derivation
from Jaynes' reinterpretation of statistical mechanics in terms of
information theory, as well as Mandelbrot's approach based on
inference theory.

\section*{Acknowledgments}

It is a pleasure to acknowledge partial financial support by FONDECYT
3080003 and 1051075. L.V. also thanks the partial financial support by the
project PNCB-16/2004 of the Cuban National Programme of Basic Sciences.

\end{document}